\documentstyle[12pt]{article}
\makeatletter
\@addtoreset{equation}{section}

\makeatother

\newcommand{\be}{\begin{equation}}
\newcommand{\ee}{\end{equation}}

\newcommand{\reff}[1]{(\ref{#1})}
\newcommand{\I}{\box{ 1\kern -2.5pt {\I}}}

\begin{document}

\title{The quark-composites approach to QCD: \\ The nucleons} 
\author{
   G. De Franceschi \\[-0.2cm]
  {\small\it Dipartmento di Fisica dell'Universit\`a  La
   Sapienza}  \\[-0.2cm]
  {\small\it and INFN -- Laboratori Nazionali di Frascati}  \\[-0.2cm]
  {\small\it I-00185 Roma, ITALIA}          \\[-0.2cm]
  \\[-0.1cm]  \and
  { F. Palumbo~\thanks{This work has been partially 
  supported by EEC under TMR contract ERB FMRX-CT96-0045}}          \\[-0.2cm]
  {\small\it INFN -- Laboratori Nazionali di Frascati}  \\[-0.2cm]
  {\small\it P.~O.~Box 13, I-00044 Frascati, ITALIA}          \\[-0.2cm]
  {\small Internet: {\tt palumbof@lnf.infn.it}}     
   }
\maketitle

\thispagestyle{empty}   

\begin{abstract}
We present a new perturbative approach to QCD based on the use of quark
composites with hadronic quantum numbers as fundamental variables. We apply it to 
the case of the nucleons by performing a nonlinear change of variables in the Berezin
integral which defines the partition function of QCD. The nucleon composites are
thereby assumed as new integration variables. We evaluate the jacobian and certain
transformation functions which appear in the change of variables.
We show that the free action of the nucleon composites is the Dirac action, and we evaluate the
first perturbative contributions to their electroweak effective action, which
turn out to be a pure renormalization. Our expansion is compatible with a perturbative as well
as nonperturbative regime of the gluons and it has the characteristic feature that 
 the confinement of the quarks is buit in.
\end{abstract}

\clearpage

\section{Introduction and summary}

"After nineteen years of study we still lack  reliable, analytic tools for
treating [the large distance region] of QCD. This remains one of the most
important, and woefully neglected, areas of theoretical particle physics"~\cite{Gros98}. We can
indeed divide the actual calculations in this field in two classes: Those which aim at
clarifying the confining mechanism, and those devoted to the description of hadronic physics,
and the difficulty stems from the fact that the former calculations are done in a framework
where it would be very awkward, if not impossible, to perform the latters. In other words, the
actual understanding of the theory requires a nonperturbative approach for low energy  and
perturbative methods for high energy, which until now it has been impossible to unify in a
unique scheme.

The direction which we take in the hope to overcome the impasse is to use quark composites with
hadronic quantum numbers as integration variables~\cite{Smit81} in Berezin integrals. This possibility has
been considered a few years ago and the necessary formalism has been developed~\cite{Palu93}. The relations
which define the physical hadronic fields in terms of the quarks cannot obviously be
inverted, and therefore the quarks cannot be eliminated. But we will see that an "effective
inversion" can be achieved in a precise way, allowing us to perform the physical calculations.
One can then hope that this will help in connection with our problem, which turns out indeed
to be the case.

The  strategy we adopt is to add to the standard action an
irrelevant operator containing the free action of the  hadronic composites and perform a
perturbative expansion by assuming  these composites as new variables of integration.
There is a fundamental difference between trilinear and bilinear composites. Surprisingly
enough for some trilinear composites, in particular the nucleon fields, the change of
variables preserves the form of the Berezin integral. As a consequence the free action of
such composites is the Dirac action.

 Also the mesonic composites can be introduced as integration 
variables, but the resulting integral does not reduce in general to a Berezin or to an
ordinary integral, and it is such that 
the propagator {\it is not the inverse of the wave operator}. It has then required 
some effort to construct an irrelevant operator containing the free action of
the pions, but this has eventually been done, along with
a first investigation of the related perturbative expansion, which has reproduced the
interactions of the chiral models.
Although identical in spirit and in its distinctive features, this expansion differs
significantly in the technicalities from that for the barions, and it is the subject of
separate works~\cite{Cara}.

Our approach is compatible with a perturbative as well as nonperturbative regime of the gluons.
Therefore, even though its applicability is  in no way restricted to a specific
regularization, we adopt a regularization on a Euclidean lattice, which is the only one
suitable for both cases, a choice which seems also natural dealing with composites. 

In this paper we consider the case of the nucleons. Our main results are the evaluation of the
jacobian and of certain transformation functions, necessary to  perform the
change of variables. Then we evaluate the first perturbative contribution to the partition
function. We expect that, since the physics of the nucleons is dominated by the exchange of
the pions, unless the fields of these particles are included, a realistic application in the
field of strong interactions is not possible. A unified treatement of nucleons and pions 
will be presented in a forthcoming paper, where the pion-nucleon interaction is derived. Here we
 restrict ourselves to the electroweak interactions of the nucleons, evaluating only
the first order terms originating from the quarks, where the pions cannot contribute. In the
presence of these interactions, the free action of the nucleons must be gauge invariant, and
it must therefore contain the appropriate electroweak couplings. The first order corrections
turn out to be a mere renormalization of the electroweak action of the nucleons. This result is
nontrivial and it provides a significant test of the consistency of our perturbative expansion.
 
 A distinctive feature of our approach is that the quark confinement is built in both for
the nucleons and the pions. This
is due to the fact that our expansion is of weak coupling for the
hadrons, but of strong coupling for the quarks. To any finite order, the quarks can
then move only by a finite number of lattice spacings, and therefore they can never be
produced in the continuum limit, so that our expansion is only compatible with quark
confinement. In this connection one can wonder whether the Wilson term to avoid spurious
quarks is necessary in our approach: Since the quarks have no poles whatsoever to finite
order, should we worry about the spurious ones ? The actual settlement of this issue requires the study of
the anomaly, which is presently under way.

To conclude a few words about the parameter of expansion. This is the inverse of a
dimensionless constant entering the definition of the nucleon composites, which is however
accompanied by the inverse of large numerical factors related to the number of quark
components, which is 24 for up and down quarks. A similar situation occurs with
the pions, but in the latter case the role of the number of quark components is 
more transparent ~\cite{Cara}, being the asymptotic parameter of a saddle point expansion. 
Since we have not evaluated any physical quantity, we cannot fix the value of the expansion
parameter, and therefore at the present stage there is nothing we can say about the rate of
convergence.

 The paper is organized as follows.
In Section 2 we report, for the convenience of the reader, the formalism for nonlinear change
of variables in Berezin integrals.

In Section 3 we define the quark composites with the quantum numbers of the nucleons,
in Section 4 we evaluate the jacobian and in Section 5 the transformation functions of the
change of variables.

In Section 6 we show that the the free action for the nucleon composites is the Dirac action.

In Section 7 we couple the nucleons to the electromagnetic field and evaluate the first
perturbative contribution to their effective action by assuming as unperturbed action the
Dirac action and the quark action $S_q$ as the perturbation. As already stated, the first
perturbative contribution, which comes from $(S_q)^3$, turns out to be simply
a renormalization  of the electromagnetic action. It would not be difficult to evaluate higher
order terms. One can indeed deduce by inspection that we will get a nucleon-nucleon
interaction, namely a fourth order term in the composites, from $(S_q)^6$ .
In such a term there can be no contribution from the gluons. Such a contribution will appear 
from $(S_q)^{12}$ in a configuration where there are nucleon composites at the
vertices of a plaquette. Whether such terms give or not a finite contribution in the
continuum limit, depends on how the  parameter $k$ scales with the lattice spacing, but again
this dependence cannot be fixed without taking into account the pions which will give
comparable or even more important contributions. 

We conclude this Section by showing that our expansion of the quark propagator is of strong
coupling.

In Section 8 we consider the weak couplings of the nucleons. Because of the well
known difficulties with chirality, our results can only be regarded as a further illustration
of the potentiality of our formalism in the generation of the structures appearing in the
composite action starting from the constituents. Again we find that the first order correction,
where the pions cannot contribute, is a pure renormalization.

\section{Barionic composites as integration variables}

We assume the convention of summation over repeated indices. 

Let us consider cubic quark composites 

\begin{equation}
\psi_I= c^{i_1,i_2,i_3}_I \lambda_{i_1}
\lambda_{i_2} \lambda_{i_3}.
\end{equation}

We want to assume these composites as new integration variables in the Berezin
integral which appears in the partition function. The above
equations define indeed a nonlinear change of variables. Of course they cannot be inverted,
but we will specify below in which sense the quark fields can be expressed in terms of the
barionic composites.

 We want to define the integral of a function of the $\psi$ in such a way that
its value be equal to that obtained by expressing these variables in terms of 
the $\lambda$, and performing the Berezin integral over the latters 
\begin{equation}
\int [d\lambda] g[\psi(\lambda)] = \int [d\psi] g(\psi).
\end{equation}
Let us restrict ourselves to functions $g$ which have a Taylor expansion. The Berezin
integral is a linear functional which associates to any function $g$ the coefficient 
 of the (unique) monomial containing all the $\lambda$ in a given order in its expansion.
Let us denote this monomial by  
\begin{equation}
\Lambda=\lambda_1\lambda_2...\lambda_{3N}.
\end{equation}
To define the integral of $g(\psi)$ over $\psi$ we must determine all the
monomials of the $\psi$ which, when expressed in terms of the $\lambda$, are
proportional to $\Lambda $ (~with nonzero coefficient). To this end we introduce the generic
monomials
\begin{equation}
   \Psi_I=\psi_1^{I_1}  \psi_2^{I_2}...\psi_N^{I_N},
\end{equation}
with degree 
\be
d_I= \sum_k I_k,
\ee
where $I$ is a vector with components $I_1,...I_N$. It is understood that if $d_I=0$, $\chi_I=1$. 
Since all the odd composites have index of nilpotency 1, $I_k=0,1$. Notice that 
\be
\Psi_I = \psi_k, \;\;if \;I_k=1,\;\;I_h=0\;\;for\; h \neq k.
\ee
We call fundamental, with weight $J$, those monomials which are proportional to $\Lambda$
 with nonzero coefficient
\be
\Psi_I =J_I \Lambda,\;\;J_I\neq 0.
\ee 

We can expand any function $g(\psi)$ in terms of fundamental monomials plus irrelevant
terms (in the sense that they do not contribute to the integral)
\begin{equation}
g= \sum_{d_I= N} g_I \Psi_I + irrelevant\;terms.
\end{equation}
The definition of the integral over the $\psi$ we are looking for is therefore
\begin{equation}
\int [d\psi] g(\psi) = \sum_{d_I= N} g_I J_I. \label{jac}
\end{equation}
Note that, although in general different expansions 
\be
g(\psi)=\sum_{d_I= N}g_I \Psi_I + irrelevant\; terms
=\sum_{d_I= N}g'_I \Psi_I + irrelevant\; terms
\ee
can exist, the above equality implies
\be
 \sum_{d_I= N}g_I J_I=\sum_{d_I= N}g'_I J_I ,
\ee
since both the lhs and the rhs are equal to the coefficient of $\Lambda$ in the
expansion of $g$ in terms of the generating elements, so that the value of the
integral does not depend on the particular expansion of $g$. 

It is remarkable that if the composites are chosen in such a way that there is only one
fundamental monomial, the integral becomes identical, apart from the weight, to the
Berezin integral over the constituents. In this case it is convenient to define the
integral over the composites exactly as a Berezin integral
\be
\int[d\psi]\Psi=1,
\ee
and regard the weight $J$ as the jacobian of the transformation. Accordingly we will 
replace the definition \reff{jac} by
\be
\int [ d \lambda ] g \left( \psi ( \lambda) \right) = J \int [d\psi] g(\psi) .
\ee 
We will restrict  ourselves to this case, which is also remarkable because more general
integrals, depending on the $\psi$ as well as the $\lambda$ can be simply evaluated
according to the equation
\be
\int [d\lambda]  g(\psi)\lambda_{i_1} \lambda_{i_2}... \lambda_{i_{3n}} =
 f_{I\; i_1i_2...i_{3n}} J \int [d\psi]  g(\psi) \Psi_I, \label{int} 
\ee
where the transformation functions $f$ are defined below. The proof follows.

By hypotesis the product of our $N$ composites is different from zero
\be
\psi_1...\psi_N = \Psi.
\ee
 We define the complementary monomials 
\be
C_I= \epsilon_I \Psi_{{\hat{I}}},
\ee
where
\be
{\hat{I}}_k = 0,1,\;\;\;for\;I_k=1,0,
\ee
and $\epsilon$ is chosen to be $ \pm1$ in such a way that the equation
\be
C_I \Psi_J = \delta_{I,J} \Psi,\;\;\;  \forall I,J, \;\;\; d_I=d_J,  \label{comp}
\ee
be satisfied.
This equation implies that 
\be
\int[d \psi] C_I \Psi_J= \delta_{I,J},\;\;\;  \forall I,J,
\ee
since, if $d_I \neq d_J$, $ C_I \Psi_J$ is never a nonzero multiple of $\Psi$ + irr.terms.
Consider now the integral
\be
{\cal I} = \int [d \lambda] g(\psi) \lambda_{i_1}...\lambda_{i_m}.
\ee
It is obviously zero if $m>3N$, the total number of quarks, but also if $m\neq 3n$, with $n$
integer. We then set, by definition 
\be
\int [d \psi] g(\psi) \lambda_{i_1}...\lambda_{i_m}=0,\;\;\;m \neq 3n.
\ee
When $m= 3n$,
\be
{\cal I}=\int [d \lambda] g^{(N-n)}(\psi) \lambda_{i_1}...\lambda_{i_{3n}},
\ee
$g^{(N-n)}(\psi) $ being the homogeneous portion of $g$ of degree $N-n$ in the expansion of
$g$ in products of the $\psi$'. Since
\be
g^{(N-n)}(\psi)= \sum_{d_I=n} g^{(N-n)}_I C_I
\ee 
with uniquely determined coefficients,
\be
{\cal I} = J \sum_{d_I=n} g^{(N-n)}_I f_{I \; i_1...i_{3n}},
\ee
which requires the following definition of the transformation functions
\be
C_I \; \lambda_{i_1}...\lambda_{i_{3n}}= f_{I \; i_1...i_{3n}}\Psi. \label{deftf}
\ee
But, of course,
\be
\int [ d \psi] g(\psi) \sum_{d_I=n} f_{I \; i_1...i_{3n}} \Psi_I=
 \sum_{d_I=n} g^{(N-n)}_I f_{I \; i_1...i_{3n}}.
\ee
Hence
\be
\int [ d \lambda] g(\psi) \lambda_{i_1}...\lambda_{i_m}=J
\int [ d \psi] g(\psi) \lambda_{i_1}...\lambda_{i_m}=
\delta_{m,3n} 
\sum_{d_I=n} f_{I \; i_1...i_{3n}} J \Psi_I
\ee
namely
\be
\lambda_{i_1}...\lambda_{i_m} \sim  \delta_{m,3n} 
\sum_{d_I=n} f_{I \; i_1...i_{3n}} \int [ d \psi] g(\psi) \Psi_I
\ee
as far as the Berezin integral is concerned.

In our physical applications the quark variables appear in two sets $\lambda_i$ and 
$ \overline{\lambda}_i$, and in the Grassmann algebra they generate we have the antilinear
conjugation $\lambda_i \rightarrow \overline{\lambda}_i$, $\overline{\lambda}_i
\rightarrow \lambda_i$ satisfying, for generic elements $\overline{\xi \eta}
= \overline{ \eta}\;\overline{\xi}$. So we have to extend the previous formulas
to include the conjugate variables. The basic Berezin integrals are
\be
\int [d \lambda] [d \overline{\lambda}]\overline{\Lambda} \Lambda=1,\;\;\;
\int [d \psi] [d \overline{\psi}]\overline{\Psi} \Psi=1
\ee
the ordering of $\overline{\Lambda}, \Lambda$ and $\overline{\Psi}, \Psi$ being
immaterial if, as it is the case in our physical applications, $N= 4\cdot$integer, since these
monomials contain an even number of quark fields. The composites $\psi_I$ are now accompanied by
the composites $\overline{\psi}_I$ obtained by conjugation.  Corresponding to Eqs. \reff{comp}
and \reff{deftf} we have by conjugation
\be
\overline{\Psi_K} \;\; \overline{C_I} = \delta_{K,I} \overline{\Psi},\;\;\; d_I=d_K,
\ee

\be
 \overline{\lambda}_{i_{3m}}...\overline{\lambda}_{i_1}\overline{C}_I=
\overline{f_{I\; i_1...i_{3m}}}\; \overline{\Psi},  \label{tf}
\ee
and from $\Psi=J \Lambda, \overline{\Psi}= J \overline{\Lambda}$, if $J$ is real, as it turns out
to be the case.
 It can then be easily checked that, if we set by definition
\be
\int [d \bar{\psi}] [d \psi]
g(\bar{\psi},\psi)\overline{\lambda}_{i_1}...\overline{\lambda}_{i_m}
\lambda_{k_1}...\lambda_{k_{m'}}=0,\;\;\; m \neq 3 \cdot \mbox{integer}, m' \neq 3 \cdot \mbox{integer},
\ee
we have
\begin{eqnarray}
 & &\int [d \bar{\lambda}][d \lambda] g(\bar{\psi},\psi)
\overline{\lambda}_{i_1}...\overline{\lambda}_{i_m}\lambda_{k_1}...\lambda_{k_{m'}}
\nonumber\\
& & =J^2 \int [d \bar{\psi}][d \psi]
g(\bar{\psi},\psi)\overline{\lambda}_{i_1}...\overline{\lambda}_{i_{m'}}
\lambda_{k_1}...\lambda_{k_{m'}} = \delta_{m,3n}\delta_{m',3n'} 
\nonumber\\
& &\sum_{d_I=n,d_K =n' }  
 \overline{f_{I\; i_{3n}...i_1}} f_{K\; k_1...k_{3n'}} J^2 \int [d \bar{\psi}] [d \psi]
g(\bar{\psi},\psi) \bar{\Psi}_I \Psi_K .
\end{eqnarray}

The conclusion is that
\be
\overline{\lambda}_{i_1}...\overline{\lambda}_{i_m} \lambda_{k_1}...\lambda_{k_{m'}}
\sim  \delta_{m,3n}\delta_{m',3n'} \sum_{d_I=n,d_K =n' } 
\overline{f_{I\; i_{3n}...i_1}}  f_{K\; k_1...k_{3n'}}\overline{\Psi}_I 
\Psi_K\;\label{tri}
\ee
is the substitution rule in the Berezin integral in the general case.

 It should now be clear in which sense we can talk of a
change of variables. Even though the constituents cannot be expressed in terms of the
composites, we only need to invert trilinear expressions in the quark and antiquark fields, and
this can be done according to Eq.\reff{tri}.

\section{The nucleon composites}

 We assume the nucleon composites to be~\cite{Ioff}
\begin{equation}
\psi_{\tau\alpha}= -{2\over3}k^{1/2}a^3
\delta_{\tau\tau_2}\epsilon_{\tau_1\tau_3} (\gamma_5
\gamma_{\mu})_{\alpha\alpha_1} (C \gamma_{\mu})_{\alpha_2\alpha_3}
(\lambda_{\tau_1\alpha_1} \lambda_{\tau_2\alpha_2} \lambda_{\tau_3\alpha_3}).
\end{equation}

In the above equation and in the following the summation over repeated indices is
understood, $C$ is the charge conjugation matrix, and  
\be
(\lambda_{\tau_1\alpha_1} \lambda_{\tau_2\alpha_2} \lambda_{\tau_3\alpha_3})=
\epsilon_{a_1a_2a_3}\lambda^{a_1}_{\tau_1\alpha_1} 
\lambda^{a_2}_{\tau_2\alpha_2} \lambda^{a_3}_{\tau_3\alpha_3}. \label{conven}
\ee
The fields $\lambda^a_{\tau \alpha}$ have color, isospin and Dirac indices $a$,
$\tau$ and $\alpha$ respectively, and are related to the up and down quarks
according to
\be
\lambda^a_{1\alpha}=u^a_{\alpha},\;\;\;\lambda^a_{2\alpha}=d^a_{\alpha}.
\ee
Correspondingly $\psi_{1\alpha},\psi_{2\alpha}$ are the proton, neutron fields.
 It is easy to check that with the above definition they transform like the quarks
under isospin and O(4) transformations.
 
Finally we should explain why we have included in the definition the cubic power of the
lattice spacing $a$ and the parameter $k$. The cubic power of a parameter with the
dimension of a length, say $l^3$, is necessary to give the nucleon field the canonical
dimension of a fermion field. At the same time a power of the lattice spacing at least cubic is
necessary to make the kinetic term irrelevant. We have written for later convenience $l^3$ in
the form $k^{1/2} a^3$, where $k$ is dimensionless and it must not diverge with vanishing lattice
spacing. 

There are altogether 8 $\psi$. Since monomials of less than 8 $\psi$ cannot obviously
contain all the $\lambda$, there can be at most one fundamental monomial $\Psi$
\be
\Psi= \psi_{1,1}... \psi_{1,4}\psi_{2,1}...\psi_{2,4}= J \Lambda.
\ee
We have chosen for the product $\Lambda$ of all the quark field components at a
given site the following ordering
\be
\Lambda=P(\lambda_{11})...P(\lambda_{14})P(\lambda_{21})...P(\lambda_{24}),
\ee
where
\be
P(\lambda_{\tau \alpha})=\lambda^1_{\tau \alpha}\lambda^2_{\tau \alpha}
\lambda^3_{\tau \alpha}.
\ee
The jacobian is obviously proportional to $k^4 a^{24}$. The nontrivial factor of
proportionality will be evaluated in the next Section.

\section{Evaluation of the jacobian}

The weight of nonrelativistic nucleon composites was evaluated in the second of
reff.~\cite{Palu93}. In this Section we consider the relativistic case. For simplicity we will
omit the trivial factor $k^4 a^{24}$ which will be reinstated at the end of the Section.

The function $\Psi$ is a homogeneous polynomial of degree 24 in the quark fields. But many
terms vanish because they contain some component of the quark field to a power higher
than 1, and the remaining terms are proportional to $\Lambda$. In order to identify
the vanishing terms and to get rid of them, we find convenient the following representation of 
the $\gamma$ matrices
\be
\gamma_k=i\left(\begin{array}{cc}
0 & \sigma_k\\
-\sigma_k & 0
\end{array} \right)
\;\;\;\;
\gamma_4=-\left(\begin{array}{cc}
0 & 1\\
1 & 0
\end{array} \right)
\;\;\;\;
\gamma_5=\left(\begin{array}{cc}
1 & 0\\
0 & -1
\end{array} \right)
\ee
and of the charge conjugation matrix
\be
C=-i\left(\begin{array}{cc}
\sigma_2 & 0\\
0 & -\sigma_2
\end{array} \right)
\ee
which satisfy the equations
\be
C^{-1}\gamma_{\mu}^T C =-\gamma_{\mu}.
\ee

In such a representation the nucleon composites take the simpler form
\begin{eqnarray}
\psi_{1i}&=&-4 \delta_{ii_2} \epsilon_{i_1i_2} (u_{+i_2}u_{-i_3}d_{-i_1})
\nonumber\\
\psi_{1i+2}&=&-4 \delta_{ii_2} \epsilon_{i_1i_2} (u_{-i_2}u_{+i_3}d_{+i_1}),\;\;\;i=1,2
\nonumber\\
\psi_{2\alpha}& = & - \psi_{1\alpha}(u \leftrightarrow d),
\end{eqnarray}
where the indices $+$ and $-$ refer to the upper and lower components of Dirac spinors.

To make the following formulae more readable we introduce the notations
\begin{eqnarray}
u^a_{+i} &=& x^a_i,\;\;\; u^a_{-i}=y^a_i
\nonumber\\
d^a_{+i} &=& z^a_i,\;\;\; d^a_{-i}=w^a_i,
\end{eqnarray}
\be
[x^a,y^b]=x^a_1y^b_2 - x^a_2y^b_1.
\ee
Therefore, if we remember the convention \reff{conven}
\be
(x_i[y,z])= \epsilon^{abc}[ x^a_iy^b_1z^c_2 - x^a_i y^b_2 z^c_1].
\ee
the nucleon composites become
\begin{eqnarray}
\psi_{1,i} &= & 4(x_i[y,w]),\;\;\;\psi_{1,i+2} =  4(y_i[x,z])
\nonumber\\
\psi_{2,i} &= & 4(z_i[y,w]),\;\;\;\psi_{2,i+2} =  4(w_i[x,z]). \label{uplow}
\end{eqnarray}
The calculation of $\Psi$ is now straightforward and it is based on the following identities
\begin{eqnarray}
(u \pi) (uAB) (uCD) &=& P(u)[(\pi C)(ABD)+(\pi D)(ABC)]
\nonumber\\
(uuA) (u \pi) & = & -2 P(u) (A \pi ),\;\;\;\pi=ab,[a,b],  \label{id}
\end{eqnarray}
$u,a,b,A,B,C,D$ being any of the quark fields $x_i,y_i,z_i,w_i$, and
\be
P(u)={1 \over 6}(xxx).
\ee
To avoid unnecessary proliferation of terms, as a guiding rule, it is convenient to 
express the $[a,b]$ in terms of $a$ and $b$ in the various $\psi$ only as far as
to be able to use the previous identities. Let $\psi_{1 1}...\psi_{1 4}= 4^4 P$,
 $\psi_{2 1}...\psi_{2 4}= 4^4 N$, so that $\Psi= 2^{16} P N$, and
\be
P=(x_1[y,w]) (x_2 [y,w]) (y_1[x,z]) (y_2[x,z]).
\ee
If we explicitate $[x,z]$, and use \reff{id} we get
\begin{eqnarray}
P &=& {1 \over 4} P(x_1) \{(y_2[y,w])(y_1z_2z_2)+c_1 \psi_{22}\} \psi_{12}
\nonumber\\
  & & - {1 \over 4} P(x_2) \{(y_2[y,w])(y_1z_1z_1)+c_2 \psi_{21}\}\psi_{11}
\nonumber\\
 &  & - {1 \over 16} \psi_{11} \psi_{12} \{(x_1y_1z_2)(x_2y_2z_1)-
(x_1y_2z_2)(x_2y_1z_1)\}. 
\end{eqnarray}
Since $ \psi_{2 \alpha}=-\psi_{1 \alpha}( x \leftrightarrow z, y \leftrightarrow w)$
it follows that
\begin{eqnarray}
N &=& {1 \over 4} P(z_1) \{(w_2[y,w]) (w_1x_2x_2)+c'_1 \psi_{12}\} \psi_{22}
\nonumber\\
  & & - {1 \over 4} P(z_2) \{(w_2[y,w]) (w_1x_1x_1)+c'_2 \psi_{11}\} \psi_{21}
\nonumber\\
 &  & - {1 \over 16} \psi_{21} \psi_{22} \{(z_1w_1x_2)(z_2w_2x_1)-
(z_1w_2x_2)(z_2w_1x_1)\}. 
\end{eqnarray}
Observe now that in the product $PN$ the terms with coefficients $c_{1,2}$,$c'_{1,2}$
do not contribute. Consider, for example, the one with coefficient $c_1$. Its factor
$\psi_{22}$, in PN, will be multiplied either with itself or with $P(z_2)$ giving zero, in
the second case, because $P(z_2)$ is the monomial of maximum degree in $z_2$, and $\psi_{22}$
is linear in $z_2$. For
this same reason it is evident that nonzero contribution can arise only from the products 
$ P(x_i)\{...\} P(z_i)\{...\},i=1,2$, (no summation), and from the product of the last two
terms in $P$ and $N$. Working out the details with the help of \reff{id}, we obtain
\be
PN= 24 P(x_1) P(x_2) P(z_1)P(z_2) R,
\ee
where
\be
R=( y_1[y,w])( y_2[y,w])(w_1[y,w])(w_2[y,w]).
\ee
Using the same procedure, explicitating, say, first the variables in the "links" with 
$y_1,y_2$, we get
\be
R= 2^3\cdot 3^2 \cdot 5 P(y_1) P(y_2) P(w_1)P(w_2),
\ee
so that, if we reinstate the factor $ k^4 a^{24}$, we get
\begin{equation}
J= k^4\; a^{24}\; 2^{22}\cdot 3^3 \cdot 5.
\end{equation}

\section{Evaluation of the transformation functions}

Also in this Section we omit the factors $k$ and $a$ which will be reinstated at the end.
According to \reff{comp} \reff{deftf} the monomial $C_{\tau \alpha}$ complementary to 
$\Psi_{\tau \alpha}$ must satisfy 
\be
C_{\tau \alpha}\psi_{\tau'\alpha'}= \delta_{\alpha \alpha'} \delta_{\tau \tau'}\Psi,
\ee
so that
\be
C_{\tau \alpha}= (-1)^{\alpha} \psi_{11}...\rlap/{\psi}_{\tau \alpha}...\psi_{24},
\ee
and the transformation functions are defined by
\begin{equation}
C_{\tau \alpha}  \lambda^{a_1}_{\tau_1 \alpha_1}\lambda^{a_2}_{\tau_2 \alpha_2}
\lambda^{a_3}_{\tau_3 \alpha_3}=f_{\tau \alpha,\tau_1 \alpha_1, \tau_2 \alpha_2,\tau_3
\alpha_3,a_1,a_2,a_3}\Psi. \label{def}
\end{equation}

As a first step to evaluate the transformation function $f$ we exploit the invariance of $\Psi$
under color, isospin and O(4) transformations. We will find in this way that $f$ is
constructed in terms of invariant tensors. Let us start by  a transformation in color space

\begin{equation}
C_{\tau \alpha}U^{a_1b_1}U^{a_2b_2}U^{a_3b_3}\lambda^{b_1}_{\tau_1 \alpha_1}
\lambda^{b_2}_{\tau_2 \alpha_2}\lambda^{b_3}_{\tau_3 \alpha_3}=
f_{\tau \alpha,\tau_1\alpha_1,\tau_2 \alpha_2,\tau_3 \alpha_3,a_1,a_2,a_3} \Psi.
\end{equation}

The above equation can be rewritten

\begin{equation}
U^{a_1b_1} U^{a_2b_2}U^{a_3b_3}f_{\tau \alpha,\tau_1 \alpha_1,\tau_2
\alpha_2, \tau_3 \alpha_3,b_1,b_2,b_3}\Psi= f_{\tau \alpha,\tau_1 \alpha_1,
\tau_2 \alpha_2,\tau_3 \alpha_3,a_1,a_2,a_3} \Psi,
\end{equation}
showing that $f$ is an invariant tensor in color, and therefore factorizes according to

\begin{equation}
f_{\tau \alpha, \tau_1 \alpha_1,\tau_2 \alpha_2,\tau_3 \alpha_3,a_1,a_2,a_3}=
\epsilon_{a_1a_2a_3}
h_{\tau\alpha,\tau_1 \alpha_1,\tau_2 \alpha_2, \tau_3 \alpha_3} . \label{ff}
\end{equation}

Let us now come to isospin and O(4) transformations. To ease the notation we represent the
pair of indices $\tau \alpha$ by the single index $i$. If under isospin,
O(4) and parity transformations, $S= \gamma_4$ 
\be
\psi'_i=S_{ij}\psi_j
\ee
then
\be
C'_i=S^c_{ij}C_j,
\ee
where $S^c$ is the contragradient representation
\be
S^c=(S^T)^{-1}.
\ee
 Performing such transformations in \reff{def} we  then find that $h$ must satisfy the
equation
\be
S^c_{kj}S_{i_1j_1}S_{i_2j_2}S_{i_3j_3}h_{j\;j_1j_2j_3}=h_{k\;i_1i_2i_3} \label{SCS} 
\ee
showing that it is an invariant tensor for the representation $S^c \otimes S^{\otimes 3}$.

For isospin transformations $S=D^{{1\over2}}$, since the quarks are in the defining
representation, and $S^c \sim S$. We must therefore look for the invariant tensors for the
representation $(D^{{1\over 2}})^{\otimes 4}$. As it is well known, the number of these
invariant tensors is equal to the number of times the identity representation is contained in
the reduction of $(D^{{1\over 2}})^{\otimes 4}$ into a sum of irreps, hence it is equal to 2.
We choose the following linearly independent tensors \be
E^{(1)}_{\tau \tau_1 \tau_2 \tau_3}=\delta_{\tau \tau_1} \epsilon_{\tau_2 \tau_3},\;\;\;
E^{(2)}_{\tau \tau_1 \tau_2 \tau_3}=\delta_{\tau \tau_2} \epsilon_{\tau_1 \tau_3}.
\ee
In the following developments we will also need the tensor
\be 
E^{(3)}_{\tau \tau_1 \tau_2 \tau_3}=\delta_{\tau \tau_3} \epsilon_{\tau_1 \tau_2}
\ee
which is related to the previous ones by
\be
E^{(1)}-E^{(2)}+E^{(3)}=0. \label{E3}
\ee
The tensor $h$ can then be decomposed according to
\be
h=  E^{(1)}\otimes t^{(1)} + E^{(2)}\otimes t.
\ee

The corresponding calculation for O(4) is  more complicated, since for the subgroup $O(4)_c$
connected to the identity, $S \sim S^c$ is the representation $ D^{\left({1\over 2},0\right)}
\oplus D^{\left(0, {1 \over 2}\right)}$, and $S^{\otimes 4}$ admits 10 invariant tensors. 
It is then convenient to determine the form of $t$ directly, rather than going trough the
intermediate step of finding these 10 invariant tensors. For this purpose we write it
in the form
\be
t_{ii_1i_2i_3}= \sum_{AB}c_{AB} (\Gamma_A)_{i_1i} (\Gamma_B C^{-1})_{i_2i_3},
\ee
where 
\be
\Gamma_A= I,\gamma_5,\gamma_5 \gamma_{\mu}, \gamma_{\mu}, \gamma_{\mu}\gamma_{\nu},
\;\;\;\mu < \nu.
\ee
No confusion should arise between the charge conjugation matrix C and the complementary
monomials $C_I$.
 Now we proceed to enforce various conditions. Firstly, as it is seen  by
inspection, $h$ must be totally symmetric under the exchange of the numbered indices. For this it
is sufficient to impose the symmetry under the permutations  $(12),(23)$
\be
(12)h=h,\;\;\;(23)h=h.
\ee
 The first of these equations yields
\be
t^{(1)}= (12)t,
\ee
which allows us to write
\be
h= (12) E^{(2)}\otimes (12)t + E^{(2)} \otimes t.
\ee
The second one, using \reff{E3}, gives two conditions on $t$
\be
(23)t=t,\;\;\;[I+(12)+(13)]t=0.  \label{23}
\ee
Since 
\be
\left(\Gamma_B C^{-1}\right)^T = \epsilon_B \Gamma_B C^{-1},
\ee
where
\begin{eqnarray}
\epsilon_B  &=& -1,\;\;\;for\;\Gamma_B=I,\gamma_5,\gamma_5 \gamma_{\mu},
\nonumber\\
\epsilon_B  &=& +1,\;\;\;for\;\Gamma_B= \gamma_{\mu}, \gamma_{\mu}\gamma_{\nu},
\;\;\;\mu < \nu,
\end{eqnarray}
the first of the above conditions implies
\be
c_{AB}=0,\;\;\;for \; \epsilon_B=-1.
\ee
Next we require $t$ to be an invariant tensor
\be
S^c \otimes S^{ \otimes 3} t=t  \label{SSS}
\ee
for S the said representation of $O(4)_c$ and $S= \gamma_4$,
namely
\be
\sum_{AB}c_{AB} S \Gamma_A S^{-1}  \otimes S \Gamma_B C^{-1} S^T=
\sum_{AB}c_{AB} \Gamma_A \otimes \Gamma_B C^{-1}.
\ee
Using the relations
\be
C^{-1}S^T = S^{-1} C^{-1},\;\;\;\;C^{-1} \gamma_4^T= - \gamma_4 C^{-1},
\ee
the above conditions reduce $t$ to the form
\be 
t = b_1 \gamma_5 \gamma_{\mu} \otimes \gamma_{\mu} C^{-1} +
b_2 \epsilon_{\mu \nu \rho \sigma} \gamma_{\mu}\gamma_{\nu} \otimes
\gamma_{\rho} \gamma_{\sigma} C^{-1}.
\ee
The second of the symmetry conditions \reff{23}, is actually implied by the first one and
\reff{SSS}, for group theoretical reasons.
To determine $t$ completely we then need two more
conditions. The first one is \be
C_{\tau \alpha} \left( \lambda_{11} \lambda_{\tau_2 \alpha_2} \lambda_{11} \right)=0.
\ee
 This follows from the observation that, in the representation where $\gamma_5$ is diagonal, 
$C_{\tau \alpha}$ contains at least 5 out of the 6 factors $\psi_{1,\alpha},
\psi_{2,i+2},\;\alpha=1,...4,\;i=1,2$. Since according to \reff{uplow} each of these 
factors is linear in $\lambda_{1,i} = x_i$, each monomial in the lhs of the above equation
contains at least 7 of these factors and must therefore vanish.
Hence
\be
h_{\tau \alpha, 11, \tau_2 \alpha_2, 11}=0
\ee
which yields
\be
b_2=0.
\ee

The second condition, which determines $b_1$ is
\be
{ 1\over 4}  \delta_{\tau \tau'} \delta_{\alpha \alpha'}= -
\delta_{ \tau'\tau_2} \epsilon_{\tau_1\tau_3}(\gamma_5 \gamma_{\mu})_{\alpha'\alpha_1}
(C\gamma_{\mu})_{\alpha_2 \alpha_3} h_{\tau \alpha \tau_1\alpha_1 \tau_2 \alpha_2 \tau_3
\alpha_3}.
\ee
 It is obtained by multiplying both sides of \reff{def} by
\be 
 - 2/3 \epsilon_{a_1a_2a_3} \delta_{\tau'\tau_2} \epsilon_{\tau_1 \tau_3}
(\gamma_5 \gamma_{\mu})_{\alpha'\alpha_1} ( C \gamma_{\mu})_{\alpha_2 \alpha_3}
\nonumber\\ 
\ee
and by summing over repeated indices. The resulting value of $b_1$ is
\be
b_1= { 1\over 96 },
\ee
so that finally, inserting the factor $a^{-3} k^{-1}$  we have
\begin{eqnarray}
h_{\tau \alpha \tau_1 \alpha_1 \tau_2 \alpha_2 \tau_3 \alpha_3}  &=& { 1\over 96}
a^{-3}k^{-1/2}  \left[\delta_{\tau \tau_2} \epsilon_{ \tau_1 \tau_3}
(\gamma_5 \gamma_{\mu})_{\alpha_1 \alpha} ( \gamma_{\mu} C^{-1})_{ \alpha_2 \alpha_3}\right.
\nonumber\\
 & & \left. + \delta_{\tau \tau_1} \epsilon_{ \tau_2 \tau_3} 
 (\gamma_5 \gamma_{\mu})_{\alpha_2\alpha}
(\gamma_{\mu} C^{-1})_{ \alpha_1 \alpha_3} \right]. \label{h}
\end{eqnarray}

\section{The free action of the nucleon composites}

Since the integral over the $\psi$ is equal to the Berezin
integral, we can assume for the nucleon the Dirac action
\begin{equation}
S_N(r_N,m_N,V)=a^4 \sum_{x y}\bar{\psi}(x) Q(r_N,m_N,V)_{x,y} \psi(y)
\end{equation}
where 
\be
Q(r_N,m_N,V)_{x,y}=-{ 1\over 2a}\sum_{\mu}(r_N-\gamma_{\mu})V_{\mu}(x)\delta_{y,x+\mu}
+\left(m_N+ {4r_N \over a}\right) \delta_{x,y}.
\ee
In the above equation $r_N$ is the Wilson parameter
\be
0 < r_N \leq 1, \label{restr}
\ee
$V_{\mu}$ is the link variable associated to the e.m. field and 
$m_N$ is the mass of the nucleon. Obviously $V_{\mu}$ and $m_N$ are diagonal
matrices in isospin space. We have adopted the standard conventions 
\begin{eqnarray}
\mu & \in & \{ -4,\ldots,4\} 
\nonumber\\
\gamma_{-\mu} & = & - \gamma_\mu 
\nonumber\\
V_{-\mu}(x) & = & V^+_{\mu}(x-\mu).\label{conv}
\end{eqnarray}

Notice that the above range of values of $\mu$ holds only for the wave operator $Q$. In the
sums occurring in the definition of the nucleon composites, and therefore of the 
transformation functions, $\mu  \in  \{ 1,\ldots,4\} $.

 Let us consider the free correlation functions
\be
<\bar{\psi}(x) \psi(y)> = {1 \over Z_N} \int [d \bar{\lambda}d\lambda]
\bar{\psi}(x) \psi(y) \exp(-S_N(r_N,m_N,1))
\ee
where the partition function 
\be
Z_N= \int [d \bar{\lambda}d\lambda]  \exp(-S_N(r_N,m_N,1)).
\ee
Introducing the $\psi$ as new integration variables
we immediately see that the nucleon composites have a canonical propagator
\be
<\bar{\psi}(x) \psi(y)>= { 1\over a^4} \left(Q(r_N,m_N,1)\right)^{-1}_{yx}.
\ee
It is perhaps worth while noticing that nothing depends at this stage on the parameter $k$
appearing in their definition, but this constant will become the inverse parameter of
expansion  when we will take into account the QCD action.

\section{ Perturbation theory and quark confinement}

In this Section we use our formalism to set up  a perturbative expansion in QCD.
Since   $S_N$  is an irrelevant  operator, it can freely be added to the standard QCD
action. We therefore assume as the total action 
\begin{equation}
S=S_N+S_G+S_q(r,m_q,Uv), \label{total}
\end{equation}
where $S_G$ is the pure gluon and em action,
\be
S_q(r,m_q,Uv)= a^4\sum_x \bar{\lambda}(x) Q(r,m_q,Uv)_{x,y}\lambda(y), 
\ee
$m_q$ is the quark mass, $r$ the Wilson parameter, which need not in general be equal 
to that of the nucleons $r_N$, but it is obviously subject to the same restriction
~\reff{restr}, $U_{\mu}$ and $v_{\mu}$ are the link variables associated to the gluon and
e.m. fields of the quarks respectively, under the conventions ~\reff{conv}.

Accordingly the partition function is

\begin{eqnarray}
Z &=&\int [dU][dv][d\bar{\lambda}d\lambda]\exp \left[-\left(S_N+S_G+S_q\right) \right] 
\nonumber\\
&=& {\cal J}\int [dU][dv][d\bar{\psi}d\psi]\exp \left[ -\left(S_N+S_G+S_q\right) \right]
 ,\;\;{\cal J}= \left(\prod_x J \right)^2.
\end{eqnarray}
In the last equality we have assumed the nucleons as new
integration variables and the quark fields must be understood their functions in the
sense specified in Section 5.
The quark action $S_q$ must now be treated as a perturbation
\be
Z={\cal J}\int[dU][dv] \exp(-S_G) \int[d\bar{\psi}d\psi]  \sum_{n=0}^{\infty} { 1\over (3n)!}
\left(-S_q\right)^{3n}\;\exp\left( -S_N \right).
\ee 
It should by now be obvious why non cubic powers of $S_q$ do not contribute. Since the 
factor $(S_q)^{3n}$ yields a factor $k^{-n}$,
because of the dependence on $k$ of the transformation functions,  we have an expansion
in inverse powers of $k$. We should emphasize that we are not treating the gauge fields
perturbatively.

We will evaluate only the first order contribution to the partition function,
where the pions cannot contribute. This splits into 2 parts
\be
- { 1\over 3!} S_q^3 = T_1 +T_2.
\ee
 $T_1$ comes from the hopping term and $T_2$ from the "mass" term. Obviously
there is no interference between the two.

Let us start from $T_1$
\begin{eqnarray}
T_1  &=& {1 \over 2^3}{ 1\over 3!}  a^9 \sum_x \sum_{\mu}
\bar{\lambda}^{a_1}_{\tau_1\alpha_1}(x) \bar{\lambda}^{a_2}_{\tau_2\alpha_2}(x)
\bar{\lambda}^{a_3}_{\tau_3\alpha_3}(x) U_{\mu}^{a_1b_1}(x) U_{\mu}^{a_2b_2}(x)
 U_{\mu}^{a_3b_3}(x) 
\nonumber\\
 & & v_{\mu,\tau_1}(x)v_{\mu,\tau_2}(x)v_{\mu,\tau_3}(x)
(r- \gamma_{\mu})_{\alpha_1\beta_1}(r-
\gamma_{\mu})_{\alpha_2\beta_2} (r- \gamma_{\mu})_{\alpha_3\beta_3} 
\nonumber\\ 
 & & \lambda^{b_1}_{\tau_1\beta_1}(x+\mu)
\lambda^{b_2}_{\tau_2\beta_2}(x+\mu) \lambda^{b_3}_{\tau_3\beta_3}(x+\mu),
\end{eqnarray}
which has been written using the fact that, to generate a nucleon field, the positions of 
the quarks must coincide with one another. Now
we will use the expressions \reff{ff} and \reff{h} for the transformation functions and the
relations
\be
\epsilon^{a_1a_2a_3} U^{a_1b_1}_{\mu}U^{a_2b_2}_{\mu}U^{a_3b_3}_{\mu} = \epsilon^{b_1b_2b_3}.
\ee
Moreover we observe that the product of the em fields acting over the quarks
generates the electromagnetic field acting over the nucleons according to
 \be
\epsilon_{\tau_2 \tau_3}v_{\mu,\tau_1}v_{\mu,\tau_2}v_{\mu,\tau_3} = 
\epsilon_{\tau_2 \tau_3} V_{\mu,\tau_1},\;\;(no\;summation).
\ee
 Since the  sum
over color indices gives a factor 6, performing also the sum over isospin indices we get
\begin{eqnarray}
T_1  & \sim & {1 \over 2^3}{ 1\over 3!}{ 12 \over 96^2} k^{-1}  a^3 \sum_x \sum_{\mu}
\bar{\psi}_{\tau \alpha}(x) V_{\mu}(x) \psi_{\tau \beta}(x+\mu)
 \left\{ 2\left( r \gamma_{\nu} \gamma_{\rho} + \gamma_{\nu} \gamma_{\mu}\gamma_{\rho}
\right)  \right.
\nonumber\\
 & &  \cdot Tr \left[\gamma_{\rho} (r + \gamma_{\mu}) \gamma_{\nu}(r - \gamma_{\mu})
\right] +
\nonumber\\
 & & \left. \gamma_{\nu} \gamma_5 (r - \gamma_{\mu}) \gamma_{\rho}(r + \gamma_{\mu})
\gamma_{\nu} (r - \gamma_{\mu}) \gamma_5 \gamma_{\rho} 
 \right\}_{ \alpha \beta}.
\end{eqnarray}
We remind the reader that the sign $\sim$ means equality under Berezin integrals.
The sums over $\nu$ and $\rho$ finally give
\be
T_1 \sim   { 3 \over 8}{ 1\over 24^2} ( 2+ r^2)  k^{-1}
a^4\sum_x \sum_{\mu} {1 \over 2a}\bar{\psi}(x)  \left( r'_N 
- \gamma_{\mu}\right)V_{\mu}(x) \psi(x+\mu), 
\ee
where
\be
r'_N= r {2 r^2 +1\over 2 +r^2}.
\ee
Notice that also $r'_N$ satisfies the restriction~\reff{restr}, and in particular $r'_N=0,1$
for$r=0,1$.

In a similar way we get
\be
T_2 \sim - { 3 \over 8}{ 1\over  24^2}  ( 2+ r^2) k^{-1}a^4 \sum_x \bar{\psi}
\left(m'_N+ { 4r'_N \over a}\right) \psi,
\ee
where
\be
m'_N = { 4 \over (2+r^2)} \left[ 2 a^2 \left( m_q + {4r\over a} \right)^3 -  r ( 2r^2+1)
{ 1\over a} \right].
\ee

In conclusion the total first order contribution can be written as a pure renormalization of
the electromagnetic action of the nucleons

\be 
\left( S_N \right)_1 = - { 3 \over 8}{ 1\over 24^2}  ( 2+ r^2) k^{-1}S_N(r'_N,m'_N,V).
\ee

Some comments concerning higher order terms.
 One can deduce by inspection that we will get a nucleon-nucleon
interaction (quartic in the composites), to second order. In such
a term there can be no contribution from the gluons. This will appear to fourth order
in a configuration where there are nucleon composites at the
vertices of a plaquette. But, as explained at the beginning, before going to higher order
we must include the pions.

We can also evaluate the quark-quark correlation function. It is obvious that at 
order n the quarks can move only by n lattice spacings. Consider for instance the quark
propagator at first order
\begin{eqnarray}
<\bar{\lambda}^a(x) \Gamma^{ab}_{xy}  \lambda^b(y)> &=& { 1\over Z} {\cal J}\int
[dU] [dv] \exp(-S_G)\int[d\bar{\psi}] [d\psi] 
\nonumber\\
 & &\bar{\lambda}^a(x) \Gamma^{ab}_{xy}  
\lambda^b(y) {1 \over 2} S_q^2 \;\exp(-S_N).
\end{eqnarray}
Notice that we have connected the quarks by a string of gluons $\Gamma$, because otherwise the
correlation function would vanish because of gauge invariance. Proceeding as in the case of
the nucleons
 we find that this correlation funtion is different from zero only for $y=x,x+\mu$. Our
perturbative series results to be a weak coupling expansion for the nucleons, but a
strong coupling expansion for the quarks. To establish quark confinement in the present
context, we should carry out this expansion to infinite order, whereas obviously we will
calculate physical processes only to finite order. Our approach, however, has the desirable
property that in such calculations quarks are never produced, a result which is only
compatible with quark confinement. It follows that in the present expansion there appear no
poles whatsoever of the quarks, and  the situation is
similar in the expansion for the pions \cite{Cara}. We therefore wonder whether we can forget
altogether the Wilson term of the quarks (but not, of course, the Wilson term of the
nucleons). A convincing assessement of this issue requires the study of the anomaly, a
problem which is under investigation.

\section{The weak interactions of the nucleons}

In this Section we study the electroweak interactions of the nucleons. Because of the well
known difficulties with chirality, however, in the present form our calculations can only
be regarded as a further illustration of the potentiality of our approach in the
reconstruction of the structures appearing in the action of barionic composites. For this 
purpose it is sufficient to restrict ourselves to the charge changing weak
interactions neglecting the Wilson term for the quarks. The corresponding action for the quarks
is \be
 S^{ch}_q =  i {1 \over 2} g \; \cos\theta \; a^4\sum_x \sum_{h=1,2}j^h_{\mu}(x)
W^h_{\mu}, \ee
where $\theta$ is the Cabibbo angle, $W_{\mu}$ are the intermediate vector bosons and
$j^h_{\mu}(x)$ is the charge changing quark current 
\be
j^h_{\mu}(x)={ 1\over 2}\left[\bar{\lambda}(x) \tau^h
\gamma_{\mu}{1 \over 2}(1-\gamma_5)\lambda(x+\mu)+
         \bar{\lambda}(x+\mu) \tau^h \gamma_{\mu}{ 1\over 2} (1-\gamma_5)\lambda(x) \right].
\ee
The above expression for the quark current is obtained by writing an action invariant under
the weak isospin and retaining the first two terms in the expansion of the related
link variables. Also the (irrelevant) action of the nucleons must be invariant under the
weak isospin transformations, so that it contains the term
\be
 S^{ch}_{\psi} = i{1 \over 2} g \; \cos\theta \; a^4\sum_x \sum_{h=1,2}J^h_{\mu}(x) W^h_{\mu},
\ee
with the corresponding charge changing nucleon current
\be
J_{\mu}^h(x) = { 1\over 2} 
\left[\bar{\psi}(x) \tau^h \gamma_{\mu} { 1\over 2} ( 1 - \gamma_5) \psi(x+\mu) +
\bar{\psi}(x+\mu) \tau^h \gamma_{\mu}{ 1\over 2} ( 1 - \gamma_5) \psi(x) \right]. 
\ee
The first order QCD correction to this action comes obviously from a term
quadratic in $S_q$
\be
\left( S^{ch}_{\psi}\right)_1 = { 1\over 2} \left(S_q(0,m_q,1)\right)^2 i {1\over 2} g
\cos\theta a^4\sum_x \sum_{h=1,2} j^h_{\mu}(x)W^h_{\mu}(x) .
\ee
 We have suppressed the em and the gluon fields for simplicity, but it
is easy to check that at this order these fields do not contribute.

Following our procedure, we find
\begin{eqnarray}
 \left(S_{\psi}^{ch}\right)_1  & \sim & {6 \over 8} { 1\over 96^2}
k^{-1} i {1 \over 2}\; g \;\cos\theta \; a^4 \sum_x \sum_{h=1,2} W^h_{\mu}(x) 
\bar{h}_{\tau\alpha,\tau_1\alpha_1\tau_2\alpha_2\tau_3\alpha_3}(\tau^h)_{\tau_3\sigma_3}
\nonumber\\
& &\;\;\;\;\; \left[\gamma_{\mu} { 1\over 2}(1 - \gamma_5)\right]_{\alpha_3\beta_3}
(\gamma_{\mu})_{\alpha_1\beta_1} (\gamma_{\mu})_{\alpha_2\beta_2}
h_{\sigma\beta\tau_1\beta_1\tau_2\beta_2\sigma_3\beta_3} \nonumber\\
& &\;\;\;\;\; { 1\over 2}
\left[\bar{\psi}_{\tau\alpha}(x) \psi_{\sigma\beta}(x+\mu) +
\bar{\psi}_{\tau\alpha}(x+\mu)  \psi_{\sigma\beta}(x) \right]. 
\end{eqnarray}

By using the expression of the structure functions and by performing the sums over isospin
and Dirac indices  we find that the first order contribution to the charge
changing weak interaction of the nucleons is a pure renormalization
\be
\left(S^{ch}_{\psi}\right)_1 \sim  - { 3 \over 8} { 1\over 24^2}  k^{-1} S^{ch}_{\psi}.
\ee

\end{document}